\documentclass[preprint,12pt]{elsarticle}

\usepackage{lineno,hyperref}
\usepackage{adjustbox}
\usepackage{threeparttable}
\usepackage{bbold}
\usepackage{xcolor}
\usepackage{multirow}
\usepackage{amsmath, amssymb}
\usepackage{cleveref}
\newcommand{\ket}[1]{\left| #1 \right\rangle} %Kinetic energy
\usepackage{url}
\modulolinenumbers[5]

\usepackage{etoolbox}
\makeatletter
    \patchcmd{\@author}{\global\let\@fnmark\@empty}{\global\let\@fnmark\@empty\global\let\@corref\@empty}{}{\@latex@error{Failed to patch \string\@author for \string\@corref reset}}
\makeatother

\newcounter{bla}

\journal{Computer Physics Communications}

\DeclareUnicodeCharacter{2212}{-}
\begin{document}

\begin{frontmatter}

\title{GPU-accelerated solutions of the nonlinear Schr\"odinger equation {\color{black} for simulating 2D spinor BECs}}

\author{Benjamin D. Smith}
\author{Logan W. Cooke}
\author{Lindsay J. LeBlanc\corref{mycorrespondingauthor}}
\ead{lindsay.leblanc@ualberta.ca}

\cortext[mycorrespondingauthor]{Corresponding author}
\address{
 {Department of Physics, University of Alberta, Edmonton AB T6G 2E1, Canada}
}

\begin{abstract}
As a first approximation beyond linearity, the nonlinear Schr\"odinger equation (NLSE) reliably describes a broad class of physical systems. 
Though numerical solutions of this model are well-established, these methods can be computationally complex.
In this paper, we {\color{black} showcase a code development approach, demonstrating} how computational time can be significantly reduced with readily available graphics processing unit (GPU) hardware and a straightforward code migration using open-source libraries.
This process shows how CPU computations with power-law scaling in computation time with grid size can be made linear using GPUs.
As a specific case study, we investigate the Gross-Pitaevskii equation, a specific version of the nonlinear Schr\"odinger model, as it describes {\color{black} in two dimensions} a trapped, interacting, two-component Bose-Einstein condensate {\color{black} (BEC)} subject to a spatially dependent interspin coupling, resulting in an analog to a spin-Hall system.  This computational approach lets us probe high-resolution spatial features -- revealing an interaction-dependent phase transition -- all in a reasonable amount of time. 
Our computational approach is particularly relevant for research groups looking to easily accelerate straightforward numerical simulation of physical phenomena.
\end{abstract}


\end{frontmatter}

% All CPiP articles must contain the following
% PROGRAM SUMMARY.

{\bf PROGRAM SUMMARY}

\begin{small}
\noindent
{\em Program Title:} \texttt{spinor-gpe} \\
{\em CPC Library link to program files:} (to be added by Technical Editor) \\
{\em Developer's repository link:} \url{https://github.com/ultracoldYEG/spinor-gpe} \\
{\em Licensing provisions(please choose one):} MIT \\
{\em Programming language:} Python                                   \\
{\em Nature of problem:}\\
Calculate the ground- and time-evolved states to the quasi-2D, pseudospinor (two-component) nonlinear Schrodinger equation, with additional Zeeman interaction and momentum-dependent, interspin coupling. \\
{\em Solution method:}\\
\texttt{spinor-gpe} is a high-level, object-oriented Python package built on Numpy and PyTorch. It propagates the pseudospinor NLSE using a time-splitting spectral method. This package accelerates solutions using NVIDIA hardware and PyTorch's cuFFT libraries and tensor functionality. \\
{\em Additional comments including restrictions and unusual features:}\\
The NVIDIA CUDA backend of PyTorch is not supported on Mac computers. To run this code on a Mac system, the dependency installation will need to exclude the CUDA toolkit. \\
\end{small}

%%%%%%%%%%%%%%%%%%%%%%%%%%%%%%%%%%%%%%%%%
\section{\label{sec:intro}Introduction}
There are many problems in physics for which the only realistic approach to a solution is through numerical techniques. The nonlinear Schr\"odinger equation (NLSE) is one such mathematical model with widespread applications throughout physics. Notably, this equation explains superfluid and magnetic properties of dilute Bose-Einstein condensates (BECs)~\cite{Radic2011,Dalfovo1999}, but it also successfully describes plasma Langmuir waves~\cite{Bao2007}, soliton dynamics~\cite{Caplan2013}, the propagation of light in nonlinear media~\cite{Kuracz2016, Brehler2017, Gong2020}, surface gravity water waves~\cite{Peregrine1983} and rogue waves~\cite{Chabalko2014}, superconductivity~\cite{Karjanto2019}, and even certain financial situations~\cite{Wroblewski2017}. The connecting thread between these disparate physical phenomena is the slowly-varying evolution of a weakly nonlinear, complex wave packet in a dispersive environment~\cite{Karjanto2019}.

A NLSE can take many forms~\cite{Silva2013}, but many physical phenomena can be adequately described with a cubic nonlinearity. For a complex scalar function $\psi$, the NLSE can be written in the general form:
\begin{align} \label{eq: NLSE}
    i \frac{\partial \psi}{\partial t} = -a \nabla^2 \psi + b\psi + c |\psi|^2 \psi,
\end{align}
\noindent where the parameter $a > 0$, and where $c$ (which can be positive or negative) represents the strength of the nonlinearity. It is worthy to note that as $c$ goes to zero, the NLSE reduces to the familiar Schr\"odinger equation. 

While very few analytical solutions of the NLSE exist, the literature is replete with techniques for finding numerical solutions~\cite{Bao2003b,Wang2007,Bao2013, Silva2013,Symes2017, Loncar2017}. Usually these methods involve direct time-integration of the equation. Throughout the last decade, these techniques have been maturing and growing more accessible, evidenced by self-contained solver packages, such as the \texttt{GPELab} toolbox for MATLAB\textsuperscript{\textregistered}~\cite{Antoine2014, Antoine2015}{\color{black}, \texttt{BEC2HPC}~\cite{Gaidamour2021}, and \texttt{GPUE}~\cite{Schloss2018}}. Despite the wealth of numerical techniques, the NLSE's nonlinear term makes it quite computationally intensive to solve~\cite{Caplan2010}. Representing small features, such as superfluid BEC vortices, often require small mesh spacings~\cite{Loncar2017}, and hence large grid sizes. 

A computer's central processing unit (CPU) operates on grid points one at a time as a ``serial processor.'' Although serial devices are optimized for low-latency, in many cases they cannot reasonably meet the demands of integrating the NLSE across large grid sizes. This results in very long computational run-times~\cite{Silva2013}, which, in extreme cases, can span days to weeks~\cite{ZamoraZamora2019}.

For certain operations and algorithms, graphics processing units (GPUs) offer a significant increase in computational power via parallelism. There are two notable types of parallel computing: (1) In \emph{task-parallelism}, akin to vehicles on an assembly line, distinct and independent sets of data are operated on concurrently. (2) In \emph{data-parallelism}, all elements of a single data collection are operated on at the same time. The Hadamard product, or element-wise matrix multiplication (represented as $A \circ B$, where $A$ and $B$ are matrices of identical size), is an example of an operation that is exceedingly data-parallel~\cite{Schloss2019}. With access to anywhere from hundreds to thousands of multiprocessors and shared memory, GPUs can leverage both types of parallelism to accelerate computations far beyond the capacity of a CPU. ``Speedup" is a relative metric for hardware performance enhancement; it is defined as  $\tau_{\rm CPU}/\tau_{\rm GPU}$, where $\tau_{\rm CPU}$ is the time per iteration on a CPU, and $\tau_{\rm GPU}$ is the time per iteration the GPU.
GPU accelerations of NLSE problems have ranged from tens~\cite{Loncar2016, Berman2015} to hundreds~\cite{Chabalko2014, Gothandaraman2011} of times. We also acknowledge that several third-party modules for GPU-accelerating the NLSE already exist~\cite{Schloss2019, Wittek2013}, however, these are not amenable to all highly-specialized research problems. 

In this paper, we introduce a general approach for GPU-accelerating numerical computations of the NLSE. Using NVIDIA graphics hardware and tools from the open-source Python community, we demonstrate how CPU-based computations with power-law scaling in grid size can scale linearly using GPU hardware.  With this paper, we also provide a packaged version of our accelerated code titled \texttt{spinor-gpe}. While this code does not claim to be generally applicable, it serves as an example of our approach for hardware-accelerating NLSE code, and as a resource to the physics community. In our particular hardware configurations, we observed $6\times$, $36\times$, and $85\times$ speedup of our pseudospinor NLSE  code. 
This required no detailed knowledge of our GPUs' architectures, and it demonstrates that a substantial computational speedup is possible using high-level programming tools like those found in the Python ecosystem.  This is particularly important for numerical calculations that make predictions for or comparisons to experimental results; rapid calculations allow for timely parameter iterations and optimizations.

This paper is organized as follows. Section \ref{sec:GPUcomp} gives a basic introduction to NVIDIA GPU operation. Section \ref{sec:theory} introduces a physically-motivated form of the NSLE and the algorithm for solving it. Section \ref{sec:implementation} describes our GPU-accelerating implementation, and Section \ref{sec:results} shows a performance comparison of our implementation across different CPU and GPU devices.  Section \ref{sec:Example} discusses the physics and results from an example calculation: a simulation of a spin-dependent gauge potential that produces quantized vortices in the spin-Hall regime. Finally, Section \ref{sec:Discussion} discusses the broader significance of our approach, before concluding with Section \ref{sec:conclusion}.

%%%%%%%%%%%%%%%%%%%%%%%%%%%%%%%%%%%%%%%%%%%%%%%
\section{\label{sec:GPUcomp}GPU Operation}
In this section, we give  a high-level introduction to GPUs, pointing out essential features and concepts, and leave details of their use in general-purpose computing to other excellent reviews~\cite{Owens2007, Owens2008}.

Graphics cards and GPUs were originally developed to render virtual 3D graphics in real-time, a task which is highly data- and task-parallel in nature~\cite{Owens2008}. While early GPUs were exceptionally good for rendering graphics, they worked with strict fixed-function pipelines. Recognizing the utility of general-purpose GPU computing, graphics card manufacturers soon developed API frameworks to directly program almost all of their GPUs' resources. There are two predominant APIs for this: OpenCL (open source, maintained by Khronos Group) and CUDA (proprietarily developed by NVIDIA Corporation). In this work, we will restrict our discussion to CUDA and NVIDIA hardware.
Similar to other frameworks, CUDA is a low-level interface to the GPU and using it requires a detailed knowledge of the GPU layout and resources~\cite{Owens2008}. Alternatively, the Python community has developed accessible packages with high-level ``pythonic'' access to back-end CUDA computing libraries, such as cuBLAS (linear algebra) and cuFFT (fast-Fourier transforms). While often used for machine learning, these packages provide a user-friendly platform for GPU-accelerating conventional scientific computations. 

In general-purpose GPU computing, one should also be aware of a device's \emph{architecture}, or the particular hardware version. Architecture includes the layout design and techniques implementing the operations, instructions, data types, registers, memory hierarchies, control units, and processors \footnote{Although companies market this as ``architecture'', this definition technically refers to a device's \emph{microarchitecture}~\cite{Hannessy2012}.} that are key factors for performance. 
Each NVIDIA device has a \emph{compute capability} (CC) metric to describe the CUDA computing features available therein. To illustrate, CC $>$ 6.x (Pascal) devices can natively perform 64-bit addition operations, whereas CC $\leq$ 5.x (Maxwell) ones cannot~\cite{CUDA2021a}.

When comparing the performance of different GPU devices, it is important to note that, \emph{within} a certain architecture, performance scales with processing core numbers, memory, and clock rates. \emph{Between} architectures, however, the vastly different hardware and instruction sets make this relationship not so simple. A rigorous performance analysis requires a detailed understanding of how the algorithm maps onto a given architecture, and is beyond the scope of this paper. Instead, benchmarking the execution time of a particular task provides a simple relative performance comparison~\cite{Silva2013, Gothandaraman2011}.

%%%%%%%%%%%%%%%%%%%%%%%%%%%%%%%%%%%%%%%%%%%%%%%%%%
\section{\label{sec:theory} Model and algorithm}
To demonstrate GPU acceleration, we investigate the Gross-Pitaevskii equation (GPE), a form of the NLSE which benefits from the GPU's features. The GPE is is used to model weakly interacting superfluids in the mean-field regime, and it is especially well-suited to describe a dilute neutral-atom BEC~\cite{Dalfovo1999}.
The GPE is well-studied in this context. Significant work has improved the path to solutions~\cite{Ruprecht1995,Adhikari2000,Bao2003b,Wang2007,Symes2017}, and illuminated a variety of physical phenomena, including vortex creation and dynamics,~\cite{Dum1998a,Jackson1998,Feder1999,Fetter2009,Radic2011,Zhang2014e,Eckel2014,Seo2016,Liu2018c} and the many-body states of spinor systems~\cite{Dutton2005,Saito2005,Zhang2005,Wang2007,De2014,Seo2016,Yukawa2020}.  Here, we take the opportunity provided by the GPU to move beyond the standard GPE: we study the physical consequences of spin- and momentum-dependent coupling, and exploit the power of the GPU to render high-resolution solutions that would otherwise be prohibitively time-expensive. 
GPU-based calculations can simulate realistic experimental conditions in a reasonable amount of time, allowing for numerically informed optimizations of experimental procedures.

\subsection{The coupled pseudospinor Gross-Pitaevskii equation}

A standard approach to studying trapped neutral-atom BECs uses the GPE, where a single-component order parameter $\psi(\textbf{r}) = \sqrt{\rho(\textbf{r})} e^{i \phi(\textbf{r})}$ represents the state of the system, where $\rho(\textbf{r})$ is  the real-space density and $\phi(\textbf{r})$ is the phase profile.  For a trapped gas of atoms, the GPE describes this order parameter as
\begin{align}
    i \hbar \frac{\partial}{\partial t}\psi(\mathbf{r}) = \left[-\frac{\hbar^2}{2m} \nabla^2 + V(\mathbf{r}) + g |\psi(\mathbf{r})|^2\right] \psi(\mathbf{r}),\label{eq:GPE}
\end{align}
where the first term in the right-hand bracket represents the kinetic energy with atomic mass $m$; the second is the trapping potential energy; and the third term is the interaction energy, where an interaction parameter $g = 4\pi\hbar a_{\text{sc}}^2/m$ is characterized by the interatomic scattering length $a_{\text{sc}}$.

Moving beyond this single-component model, we next consider the \emph{spinor} condensate: a two- (or more-) component system where a higher-dimensional order parameter describes the density of atoms in two (or more) spin states. In describing experimental systems with alkali metal atoms, these spin states are pseudospins whose real identities are defined by $m_F$ levels in the ground state manifold.  In the case of a two-spinor, or pseudospinor, the order parameter takes the form $\psi \rightarrow {\color{black}\Psi} = \{\Psi_{\uparrow},\Psi_{\downarrow}\}$.  
The interaction terms of the spinor GPE model must also account for the possibility of distinct inter- and intra-spin scattering lengths. Indeed, pseudospinors have three unique interaction strengths: $g_{\uparrow\uparrow}$, $g_{\downarrow\downarrow}$, and $g_{\uparrow\downarrow}=g_{\downarrow\uparrow} \equiv g_\updownarrow$.

Next, we introduce external fields that are carefully chosen to give a spin-dependent coupling between the pseudospins. These yield ``artificial gauge fields''~\cite{Spielman2009, Dalibard2011} that can mimic the effects of magnetic fields, electric fields, and/or spin-momentum coupling in these atomic systems.
We consider here the case where two lasers with opposite propagation directions ($\pm\hat{x}$) effect a two-photon Raman transition, thereby producing a spatially-periodic spin-wave in the BEC along the $\hat{x}$ recoil direction\footnote{In this work, we limit the discussion to a spin wave, and thus momentum transfer, along one dimension, but this type of interaction can be extended to additional dimensions~\cite{Wu2016,Anderson2012}}.
This spatial-periodicity can be removed via a unitary transformation, resulting in the two bare spin dispersion relations shifting opposite directions in $\hat{k}_x$ momentum space. In this rotated picture, this system is described by the effective single-particle energy Hamiltonian~\cite{Lin2011, Radic2011}:
\begin{align}
    \label{eq:RamanSingle}
    \hat{\mathcal{H}} = \left[\frac{\hbar^2 \mathbf{k}^2}{2m} + V(\mathbf{r}) \right] \check{\mathbb{1}} - \frac{\hbar^2 k_{\rm L} \hat{k}_x}{m}\check{\sigma}_z + \frac{\hbar\Omega(\mathbf{r})}{2}\check{\sigma}_x + \frac{\hbar\delta(\mathbf{r})}{2}\check{\sigma}_z,
\end{align}
 where $k_{\rm L}$ is the magnitude of the lasers' wavevector, $\delta(\mathbf{r})$ is the two-photon Raman detuning
 and $\{\check{\mathbb{1}}, \check\sigma_x, \check\sigma_y, \check\sigma_z \}$ are the identity and Pauli matrices in the spinor basis. {\color{black} The Raman coupling between the ground states is generated via two electric fields having an equally far detuning $\Delta_e$ from some excited level. In this regime, the excited state can be adiabatically eliminated, and the total coupling strength takes the form $\Omega(\mathbf{r}) = \Omega_1(\mathbf{r}) \Omega_2(\mathbf{r}) / 2\Delta_e$, where the $\Omega_i$ are the Rabi frequencies, a measure of the electric dipole couplings between the excited and individual ground states~\cite{Steck2007}.} The characteristic energy scale of this Hamiltonian is $E_{\rm L} = \hbar^2 k_{\rm L}^2 / 2m$, the kinetic energy imparted to an atom by a single-photon recoil. Experimentally, spatial dependence in the detuning and Raman coupling can readily be achieved with a spatially-dependent magnetic field (via the Zeeman effect) or a spatially-varying optical intensity (via the ac Stark effect), respectively. Spatial-dependence in the optical field can be obtained using, for example, a spatial light modulator device.

Finally, we incorporate the last three terms of Eq.~(\ref{eq:RamanSingle}) into the  GPE pseudospinor Hamiltonian Eq.~(\ref{eq:GPE}), taking into account the appropriate signs of the detuning and the momentum shift. We interpret the spinor components $\{\Psi_{\uparrow}(\mathbf{r}),\Psi_{\downarrow}(\mathbf{r})\}$ as bare spins that have undergone a spin-dependent momentum shift $\{|\uparrow, -k_L\rangle, |\downarrow, +k_L\rangle\}$~\cite{Spielman2009}. After converting all quantities to dimensionless ones (denoted by tildes), the {\color{black} NLSE equation describing $\widetilde{\Psi}$ is}
 
{\color{black}
\begin{align}
&-i \frac{\partial}{\partial \tilde{t}} \widetilde\Psi = \left[\mathcal{H}^{(1)} + 
\mathcal{H}^{(2)} + \mathcal{H}^{(3)} \right]  \widetilde\Psi +\mathcal{H}^{(4)} \widetilde\Psi, \label{eq:GPEcomponent}\\
\intertext{where}
&\mathcal{H}^{(1)} = -\tfrac{1}{2}{\tilde{\mathbf{k}}^2}\check{\mathbb{1}}  +  \tilde k_{\rm L}\tilde k_x \check{\sigma}_z
\label{eq:H1}\\
\label{eq:H2}
&\mathcal{H}^{(2)} =  \widetilde{V}(\tilde{\textbf{r}})\check{\mathbb{1}} +  \tfrac{1}{2}{ \tilde \delta(\tilde{\textbf{r}})}\check{\sigma}_z \\
\label{eq:H3}
&\mathcal{H}^{(3)} = 
\begin{pmatrix}
g_{\uparrow\uparrow} |\tilde{\Psi}_\uparrow(\tilde{\mathbf{r}})|^2 + g_\updownarrow |\tilde{\Psi}_\downarrow(\tilde{\mathbf{r}})|^2 & 0 \\
0 & g_{\downarrow\downarrow} |\tilde{\Psi}_\downarrow(\tilde{\mathbf{r}})|^2 + g_\updownarrow |\tilde{\Psi}_\uparrow(\tilde{\mathbf{r}})|^2
\end{pmatrix} \\
\label{eq:H4}
&\mathcal{H}^{(4)} = \tfrac{1}{2}{ \tilde \Omega(\widetilde{\textbf{r}})} \check{\sigma}_x
\end{align}}
represent the kinetic [Eq.~(\ref{eq:H1})], potential [Eq.~(\ref{eq:H2})], interaction [Eq.~(\ref{eq:H3})], and the Raman coupling [Eq.~(\ref{eq:H4})] energies.
%; the upper (lower) signs in these equations refers to the $\uparrow (\downarrow)$ component.
This pseudospinor GPE describes the emergence of both superfluid effects (e.g. quantized vortices~\cite{Radic2011}) as well as magnetic structures (e.g. stripes, spin domains~\cite{De2014}).

\begin{figure}[tb!]
    \centering
    \includegraphics[width=85mm]{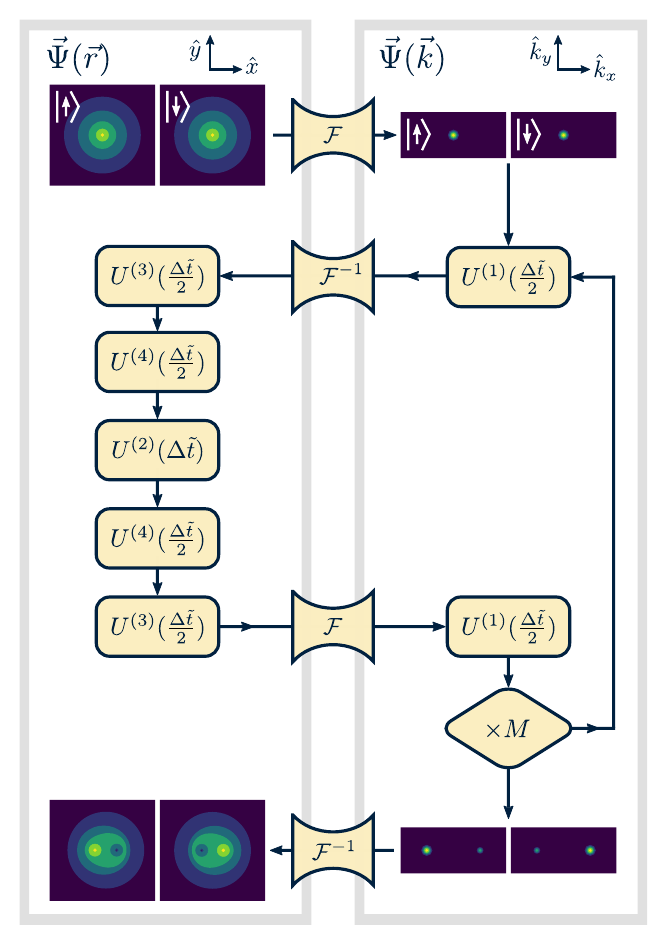}
    \caption{A flow diagram of the time-splitting spectral algorithm used in this paper. Starting from some initial order parameter $\vec{\Psi}(\tilde{\textbf{r}})$, we propagate in time by applying evolution operators $U_i$ [see Eqs. (\ref{eq:H1})-(\ref{eq:TEOp})] and fast Fourier transforms $\mathcal{F}$. The operations applied sequentially in the loop represent those for a single propagation time step $\Delta \tilde{t}$. In our Python code, the operations for a single time step are organized into a function that is called in a loop $M$ times. After the loop terminates, the final order parameter is available.}
    \label{fig:Algorithm}
\end{figure}

\subsection{\label{ssec:Algorithm}Algorithm}
In this section, we describe the basic elements of our algorithm for integrating solutions to the GPE of Eq. (\ref{eq:GPEcomponent}). We begin by assuming that the BEC is confined to the $x-y$ plane by a strong harmonic trapping potential in the transverse direction with frequency $\omega_z$. Out-of-plane excitations are suppressed, and the dimensionless order parameter along this dimension takes on a Gaussian profile with unit norm. If the energy level spacing $\hbar \omega_z$ is larger than the interaction energy, we can approximate our system as quasi-2D~\cite{Radic2011, Smith2005} and represent the order parameters and operators on 2D grids. 

Our method for integrating solutions to the pseudospinor GPE relies on the well-established time splitting spectral (TSSP) method~\cite{Bao2002} (see Figure \ref{fig:Algorithm}). {\color{black} Solutions are evenly discretized in space on $N_x \times N_y$ grids. The real-space grid extends over $(\pm x^{\rm max}, \pm y^{\rm max})$, with spacings $(\Delta x, \Delta y)$; the momentum space grid extends over $(\pm \pi / \Delta x, \pm \pi / \Delta y)$ with spacings $(\Delta k_x, \Delta k_y) = (\pi / x^{\rm max}, \pi / y^{\rm max})$. The solutions} are propagated through time by repeatedly applying time-evolution operators to the previous time step's order parameter. The time-evolution operators are given by
\begin{align} \label{eq:TEOp}
U^{(n)} (\Delta \tilde t) = \mathrm{exp}(i \mathcal{H}^{(n)} \;\Delta \tilde t),
\end{align}
where $\Delta \tilde t$ is a unitless {\color{black} discrete} time step smaller than any relevant time scales of the system. Propagation in real-time yields the dynamics of the spinor system, while propagation in imaginary-time ($\Delta \tilde t \rightarrow i \Delta \tilde \tau$) asymptotically approaches ground state solutions. ``Applying’’ an evolution operator amounts to Hadamard multiplication of the complex operator array and the order parameter array. 
In the TSSP method, a Fourier transform takes the order parameter to momentum space where $\mathcal{H}^{(1)}$ is diagonal, i.e. $\nabla^{2} \rightarrow -\mathbf{\tilde k}^2$.
In this way, we perform two 2D FFTs to avoid the more computationally-expensive finite-difference Laplacian in the $U^{(1)}$ operator. 

Starting from an initial spinor order parameter, we propagate the GPE in a loop over $M$ discrete time steps of length $\Delta \tilde t$. We pre-compute and store the evolution operators [Eq. (\ref{eq:TEOp})] corresponding to the potential {\color{black} $\tilde V(\tilde{\textbf{r}}) \check{\mathbb{1}}$}, kinetic $\mathcal{H}^{(1)}$, Raman coupling {\color{black}$ \mathcal{H}^{(4)}$}, and Raman detuning {\color{black} $\tilde \delta(\tilde{\textbf{r}}) \check{\sigma}_z$} energy components, since they are constant throughout the propagation loop; the nonlinear mean-field terms $\tilde g_{ij}|\widetilde\Psi_i(\tilde{\mathbf{r}})|^2$ depend on the densities, and therefore are calculated at each time step. As shown in Figure \ref{fig:Algorithm}, we apply the $U^{(1)}$, $U^{(3)}$, and $U^{(4)}$ operators with the familiar Strang splitting for stability and to reduce errors induced by the various non-commuting $U^{(i)}$ operators~\cite{Bao2013}{\color{black}; because the 2D operators additionally live in spin space, the real-space evolution operators $U^{(2, 3, 4)}$ do not necessarily commute and should be split to second-order
~\cite{Bao2004, Bao2005}.} {\color{black} Previous demonstrations using higher-order splitting have further improved spatial accuracy, but the form of these splittings as applied to spinor systems was not straightforward~\cite{Bao2005}}. Within a single time-step loop, four 2D FFTs and $\gtrsim 20$ Hadamard products are performed. Since probability density is not conserved in imaginary-time propagation, we normalize the order parameters to the total atom number at each time step.

While the individual operations are highly data-parallel, they must be applied sequentially. Hence in our case, it was advantageous to maintain data on a single device, avoiding the additional transfer times between various devices in so-called ``heterogenous'' or distributed computing configurations~\cite{Silva2013}. This way, the data does not leave the device until the entire simulation is complete.

\section{\label{sec:implementation} Implementing GPU acceleration}
This section describes how we adapted our previously-existing simulation code for GPU acceleration. Our original (non-GPU-compatible)  implementation of the GPE exclusively employed the NumPy scientific computing library.
The algorithm relied heavily on the FFT and Hadamard product, both of which have the potential to be highly data-parallel operations. In search of accelerated execution times, we considered both the software and hardware aspects of our computations.

Several established CUDA-compatible Python packages, such as Tensorflow, provide wrappers of the needed cuFFT library{\color{black}~\cite{CUDA2021b}}. However, we settled on PyTorch, a relatively newer package, because it has a ``native Python” interface and is intentionally designed to have similar, if not identical, syntax to NumPy. This implies a short learning curve and minimal changes to our original code~\cite{Harris2020}; most of the changes we made to our code while migrating were drop-in replacements. As with other machine learning packages, PyTorch code can execute on either a CPU or a CUDA-enabled GPU, with a simple software switch between the two; this made it convenient to develop and test our code on a CPU before scaling it up to run on a GPU workstation. Our GPU code also takes advantage of the complex data type recently released for PyTorch.

\begin{table}[tb!]
    \centering
    \begin{tabular}{l | c c c}
        \hline
        \multicolumn{1}{c|}{\multirow{2}{*}{GPUs}} & GeForce & GeForce & TITAN \\ 
         & MX150 &  980 Ti & V\\\hline
        Architecture & Pascal & Maxwell & Volta\\ 
        Compute Capability & 6.1 & 5.2 & 7.0\\ 
        \# CUDA Cores & 384 & 2816 & 5120\\ 
        Clock (Boost) [GHz] & 1.47 (1.53) & 1.0 (1.07) & 1.2 (1.45)\\ 
        VRAM Mem. [GB] & 2.0 & 6.0 & 12.0\\ 
        Mem. Bandwidth [Gbps] & 48.06 & 336.6 & 651.3\\ 
        Mem. bus width [bits] & 64 & 384 & 3072\\ \hline
        \multicolumn{4}{c}{} \\ \hline
        \multicolumn{1}{c|}{\multirow{2}{*}{CPUs}} & Intel i5- & AMD FX- & Intel i9- \\ 
         & 7200U & 6300 & 9900K \\\hline
        Clock (Boost) [GHz] & 2.5 (3.1) & 3.5 (4.1) & 3.7 (5.0)\\ 
        Available RAM [GB] & 8 & 16 & 32\\ 
        \hline
    \end{tabular}
    
    \caption{Our PyTorch implementation can execute on any of our CUDA-enabled NVIDIA graphics cards (top) and our CPUs (bottom). The corresponding GPU/CPU hardware pairs are installed on a commercial laptop and two custom-built workstations, respectively. Key specifications of these devices are given. While RAM is not a property of CPUs, it's included here as a computational resource for the devices.}
    \label{tab:specs}
\end{table}

On the hardware side, we constructed two different computer workstations with NVIDIA graphics cards. Our first workstation contains a GeForce 980 Ti (Maxwell arch., C.C. 5.2), a common commercial gaming graphics card. Our second workstation contains a Titan V (Volta arch., C.C. 7.0). In addition to the two workstations, we also had a commercial Acer Aspire laptop with an integrated NVIDIA GeForce MX150 graphics card (Pascal arch, C.C. 6.1). Specifications for these three devices, along with their corresponding CPUs, are summarized in Table \ref{tab:specs}.

%%%%%%%%%%%%%%%%%%%%%%%%%%%%%%%%%%%%%%%%%%%%%%%%%%%%%

\section{\label{sec:results}Performance Benchmarking}
In this section, we show and analyze the benchmark results of timing our propagation stepping function (see Figure \ref{fig:Algorithm}, caption): we compare the performance and scaling of our three GPUs and three CPUs with increasing grid sizes. Note that these benchmarks only compare the performance of our PyTorch code on different devices, one device at a time. 

To make a fair comparison between GPU and CPU performance, we transferred the order parameter and energy grids to the GPU’s memory before running the benchmarks on those devices, thereby avoiding the relatively slow data transfer rate between computer RAM and GPU VRAM~\cite{Chabalko2014}. We exclusively employ complex double-precision floats (\texttt{torch.complex128}) in our simulations and benchmarks. 

\begin{figure}[tb!]
    \centering
    \includegraphics[width=\columnwidth]{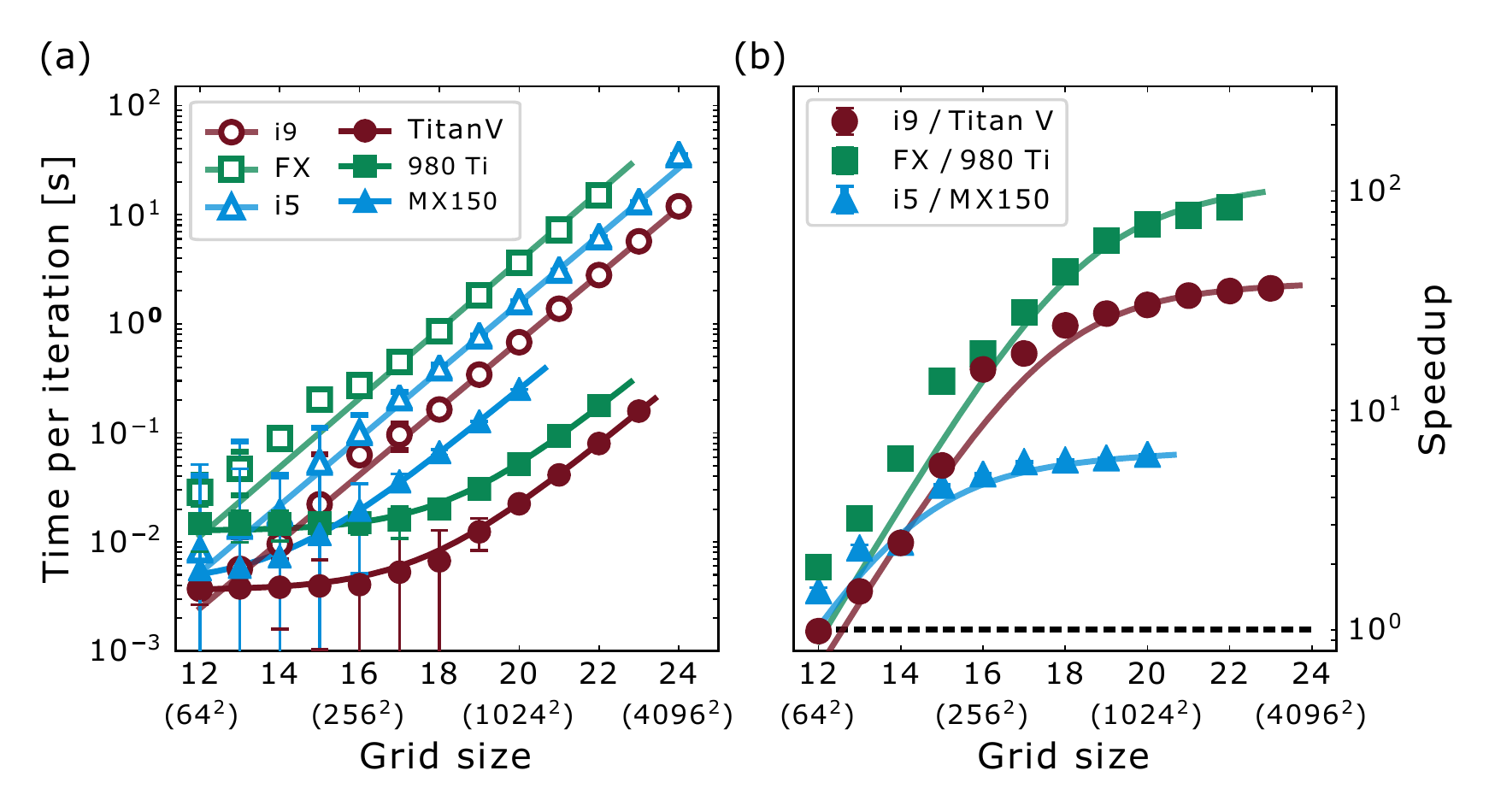}
    \caption{(a) Performance comparison benchmark of the propagation step function for the six devices across various grid sizes, shown here on a log-10/log-2 plot. The points represent the median evaluation time of many trials, and the error bars represent the median absolute deviation. The CPU and GPU evaluation times were fit by power law [open symbols, {\color{black}$\kappa=1.03(3)$}] and linear [filled symbols, {\color{black}$\theta = 9.7(1) \times 10^{-8}~\text{s}/N$}] models, respectively. (b) Speedup of the hardware pairs as a function of grid size. The black dashed line represents the break-even performance for the hardware pairs. {\color{black}For comparison and reference, the files used to generate these data are found in the repository subdirectory \href{https://github.com/ultracoldYEG/spinor-gpe/tree/master/spinor\_gpe/benchmarks}{/spinor\_gpe/benchmarks}.}}
    \label{fig:Bench}
\end{figure}

We measured the propagation function evaluation times using the Python \texttt{timeit} module. We separately timed many different evaluations of our propagation stepping function, and repeated this process on each device for various 2D grid sizes $N = N_x \times N_y = 2^\eta$, where $\eta \in \mathbb{N}$ and $N$ is the size of a single spinor component. Due to concurrent system processes that we could not eliminate, the distributions of evaluation times were highly non-Gaussian. Although there exist sophisticated benchmark analysis techniques for understanding these types of distributions~\cite{Chen2015, Hoefler2015}, the median and median absolute deviation provided a simple and interpretable statistic and uncertainty for our purposes. With each GPU device, there was a maximum grid size above which the data could no longer fit into VRAM, and the benchmark halted. The results of evaluation time versus $\eta$ are plotted in Figure \ref{fig:Bench}(a), and the speedup between particular CPU/GPU pairs is given in Figure \ref{fig:Bench}(b). At the smallest grid sizes $(\eta = 12)$, the performance of the GPUs and CPUs were comparable. The largest speedups we measured for each device pair (from smallest to largest) were 6.3, 36, and 85. 

\begin{figure}[tb!]
    \centering
    \includegraphics[width=\columnwidth]{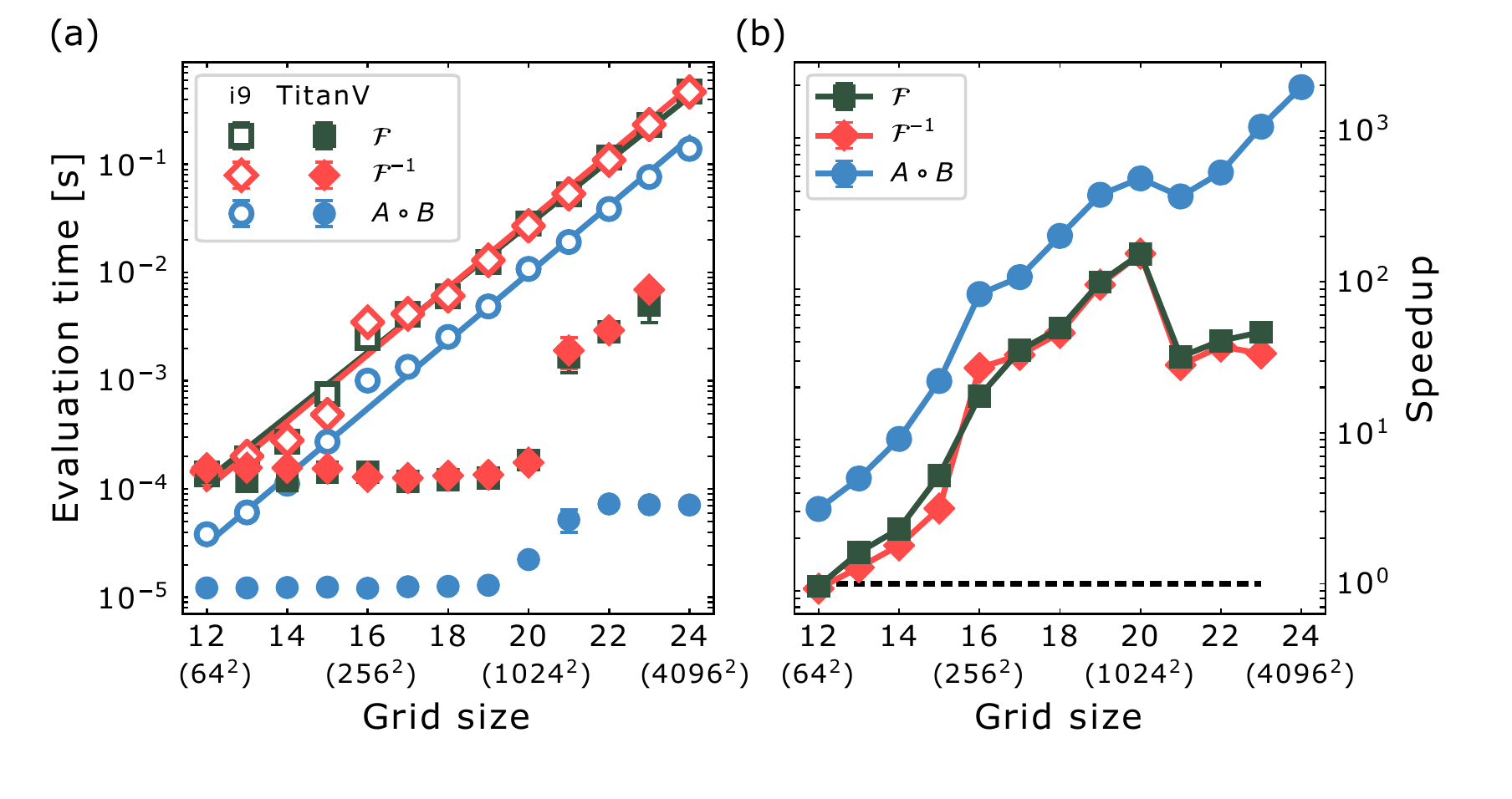}
    \caption{(a) The evaluation times for individual forward and inverse 2D FFT and Hadamard product $(A \circ B)$  function calls over various grid sizes, computed on the i9 CPU (open symbols) and the Titan V graphics card (filled symbols). The points represent the median evaluation time over many trials, and the error bars represent the median absolute deviation. The CPU function times are well-described by a power-law, with an average exponent of {\color{black}$\kappa = 1.0154(2)$}. (b) The speedup of the three function evaluation times as a function of grid size.}
    \label{fig:Bench_Jump}
\end{figure}

The CPU evaluation time data are well-described by a power-law function of the form {\color{black}$f_C(N) = \lambda N^\kappa$}. The three CPU evaluation times scale vary similarly, and, on average, {\color{black}$\kappa = 1.03(3)$}. The GPUs, in contrast, scaled linearly as {\color{black}$f(N) = \theta N + \xi$}, with {\color{black}$\theta = 9.7(1) \times 10^{-8}~\text{s}/N$} and a device-dependent offset. The scaling difference between the two device classes highlights the power of GPU data parallelism. This linear scaling means that doubling of smaller grid sizes makes negligible difference to the evaluation times. Our GPUs basically provided increased resolution for free.

As mentioned previously, it is generally difficult to interpret benchmarks for GPUs from different device architectures. There are some insights, however, that we can gain from testing the performance of the atomistic 2D FFT and Hadamard functions that compose our algorithm. Figure \ref{fig:Bench_Jump}(a) shows evaluation times for each of these functions on the i9 CPU and the Titan V GPU. The times measured on the i9 all display a power-law scaling with increasing grid size. The Titan V data show an entirely different behavior: at small grid sizes, the Hadamard and FFT times are all constant, but begin to rise near $\eta = 19$ to $20$. We interpret this point of change as the Titan V's limit for simultaneous data operations; at grid sizes larger than this, the device must batch the data and operate on those batches sequentially. The improved scaling of the functions' evaluation time by the Titan V device supports the use of GPU computing for our algorithm.

\section{\label{sec:Example}Simulation Example: Spin Hall System}

In this section, we demonstrate our GPU-accelerating method by simulating the ground states of a spin Hall system. In such a system, the spin-up and spin-down constituents experience effective magnetic fields of equal magnitude but opposite direction.  The spin Hall effect has been investigated theoretically~\cite{Zhu2006, Liu2007} and experimentally~\cite{Beeler2013} using Raman-induced spin-orbit coupling in ultracold atoms; the presence of a spatial gradient in the Raman coupling and an effective ``electric'' force (a role played by gravity) generate transverse spin Hall currents. Other work showed that interspin interactions can greatly alter the properties of spin Hall states~\cite{Furukawa2017, Furukawa2014}.
In the simulations that follow, we investigate the mean-field ground states of a two-component BEC subject to a spatially varying spin-dependent gauge potential, considering various interspin interaction strengths. Similar to the proposal given in \cite{Beeler2013}, these states are generated \emph{in situ} in the absence of any effective electric force and reside in the classical spin Hall regime ($\nu = N / N_\phi \gg 1$)~\cite{Furukawa2017}. 

We consider a harmonically confined, pseudospinor BEC of $10^4$ atoms, where the trapping frequencies $\omega_z \gg \omega_x = \omega_y \equiv \omega_{\perp}$. The harmonic oscillator length $a_0 = \sqrt{\hbar / m \omega_{\perp}}$  and energy $E=\hbar\omega_{\perp}$ set the characteristic length and energy scales of our system. The intraspin interaction parameters are  $g_{\uparrow\uparrow} = g_{\downarrow\downarrow} \equiv g$.

The single-particle physics of this problem  can be analysed by diagonalizing the Hamiltonian [Eq.~(\ref{eq:RamanSingle})].  For weak coupling ($0 < \hbar\Omega < 4 E_{\rm L}$), the lower-energy band  of the dispersion relation takes on a double-well shape~\cite{Lin2011}; as $\Omega \rightarrow 0$, the two minima reside at $\pm k_{\rm L}$. The eigenstates in this band vary across $k$, with $\ket{\downarrow}$ dominating the state near $+ k_{\rm L}$, and $\ket{\uparrow}$ dominating the state near $- k_{\rm L}$ (even for nonzero $\Omega$). If the atoms are confined to the lowest energy band, this amounts to a spin-dependent Abelian gauge potential, and the following effective Hamiltonian applies: 
\begin{align}
    \hat{\mathcal{H}}_{\rm eff,x} = \frac{\hbar^2}{2m}\left(k_x + \mathcal{A}_x^*\check{\sigma}_z\right)^2,
\end{align}
where the magnitude of the artificial gauge potential $\mathcal{A}_x^*$ multiplies the Pauli matrix in the dressed-spin basis, and scales as 
\begin{align}
    \mathcal{A}_x^* = k_{\rm L} \left[1 - \left(\frac{\hbar\Omega}{4E_{\rm L}}\right)^2\right]^{1/2}
\end{align}
for  $\hbar\Omega \leq 4E_{\rm L}$ and $\delta = 0$~\cite{Lin2011}.
%Since the dressed spins are nearly equivalent to the bare spins near the minima of the dispersion,
A spatially-varying $\Omega(y)$, and hence $\mathcal{A}_x^*(y)$, produces a synthetic magnetic field for each spin
\begin{align}
    B_{\uparrow(\downarrow)}^* \hat{z} = \pm \left(\frac{\hbar}{q}\right) \nabla \times \mathcal{A}_x^*(y)
\end{align}
that is equal in magnitude, but opposite in direction. 

\begin{figure*}[tb]
\includegraphics[width=\textwidth]{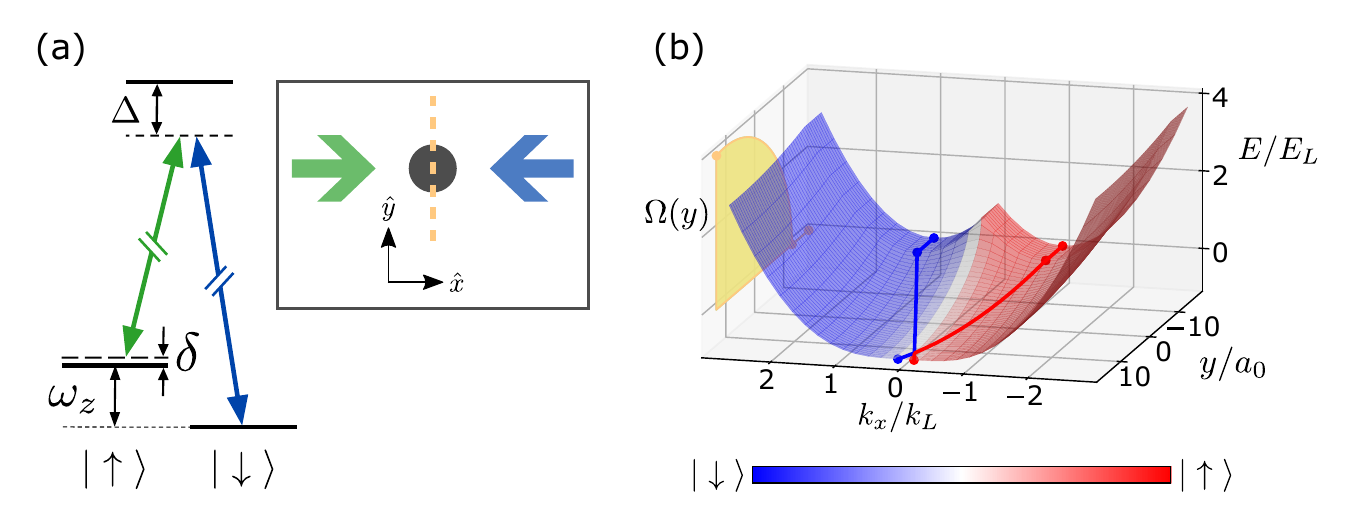}

\caption{\label{fig:simsetup} (a) A pair of counter-propagating Raman lasers couple two atomic levels of a harmonically-confined BEC. A spatial light modulator device (not shown) tailors the laser intensity to vary proportional to $\Omega(y)$ [Eq.~(\ref{eq:Omega})] along the yellow dashed line. (b) The Raman momentum-energy dispersion as a function of $y$-position due to the spatially-tailored Raman coupling profile shown in on the left 3D wall. This coupling is assumed to be uniform along the $x$-direction. The solid red and blue curves indicate the double-well minima, or the gauge potentials $\mathcal{A}_x^*(y)\hat{\sigma}_z$.}
\end{figure*}

We imposed a spatially-varying Raman coupling profile of the form
\begin{align} \label{eq:Omega}
    \frac{\hbar\Omega(y)}{E_{\rm L}} = \sqrt{8y-y^2}
\end{align} 
to linearize $\mathcal{A}_x^*(y)$. As shown in Figure \ref{fig:simsetup}(b), this created two degenerate spin-dependent wells in $k$-space that moved inward from $k_x = \pm 1 \rightarrow 0$ as $y$ increased. The induced uniform, synthetic spin-dependent magnetic field had a magnitude of $|B_{\uparrow(\downarrow)}^*| = 0.369 ~\hbar/q a_0^2$ across the region of the BEC. Before each simulation trial, we seeded each order-parameter component with 50 randomly-placed vortices having a spin-dependent winding. Due to the spin-dependent nature of the magnetic field, we expected the components to acquire opposite angular momentum. We then propagated the solution in imaginary time over two segments of 60,000 steps each. In the first segment, we periodically ``annealed’’ the system with Gaussian noise to more quickly find the ground state; second, we propagated without annealing. Good convergence to the ground state could typically be obtained following this method.

\begin{figure*}[tbp]
\includegraphics[width=\textwidth]{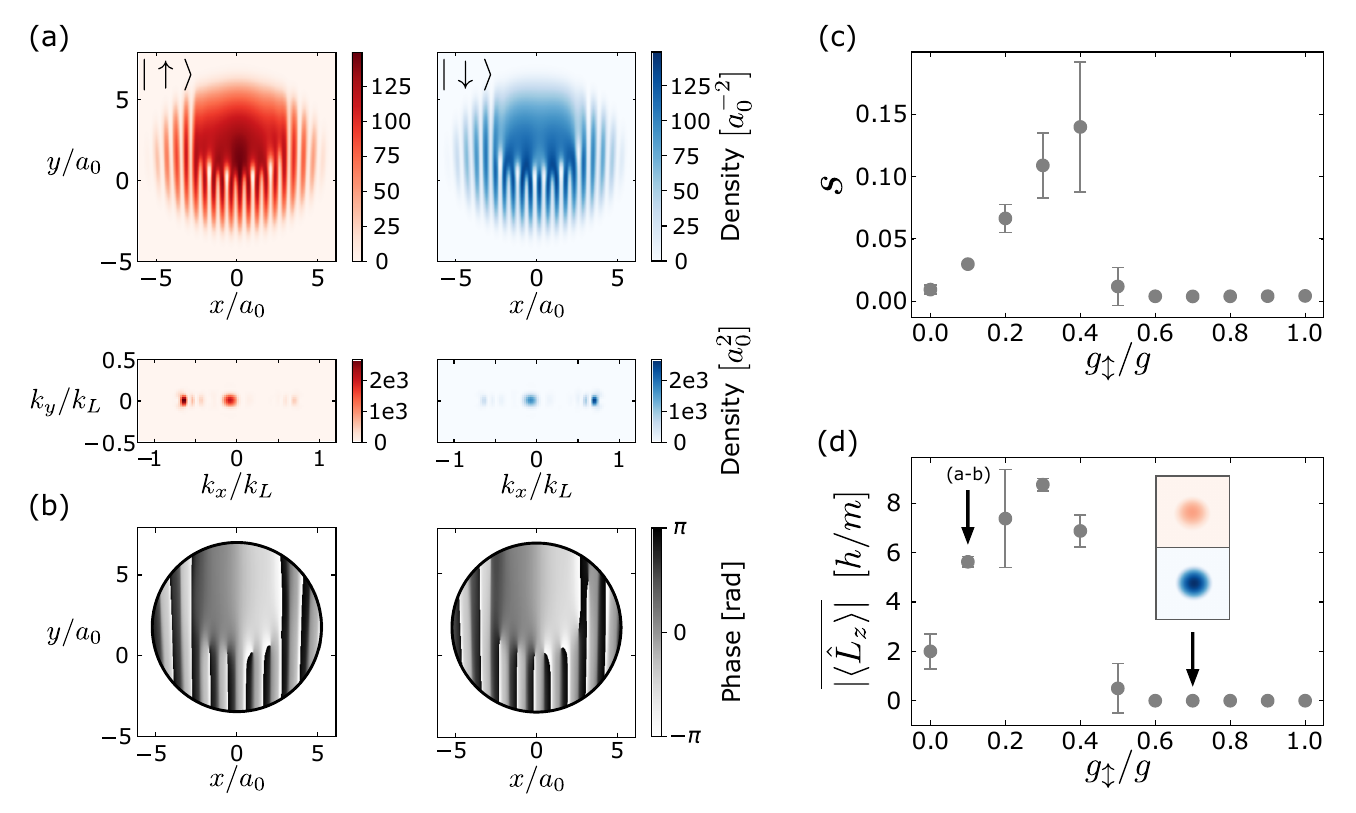}

\caption{\label{fig:simresults} (a) Enlarged detail of the calculated real-space (top, 2.6$\times$ mag.) and k-space (bottom, 9.8$\times$ mag.) ground state densities on a 1024 x 1024 grid for $g_{\updownarrow} = 0.1$. The trapping frequencies are $(\omega_{\perp}, \omega_z) / 2\pi = (50, 2000)$ Hz. The smooth central regions of each real-space density correspond to momentum components near $k_x = 0$. (b) The spatial phase profiles $\phi_{\uparrow(\downarrow)}(x, y)$ of the solutions from (a), showing opposite vortex windings in each component. (c) The phase separation of the two components as a function of the interspin interaction strength $g_{\updownarrow} / g$. Error bars indicate the standard deviation of several trials. (d) The absolute value of $\langle \hat{L}_z\rangle_{\uparrow(\downarrow)}$ averaged for both components, as a function of $g_{\updownarrow} / g$. We expected that the circulation, and hence the magnitude of the $B_{\uparrow(\downarrow)}^*$, experienced by each spin component would be the same, however, interactions and the initial random seeding generally tended to imbalance the respective angular momenta for a given simulation trial. For values of $g_{\updownarrow} / g$ larger than $\sim 0.5$, the spins were completely phase-mixed with no angular momentum present in either component.}
\end{figure*}

We simulated and characterized ground state solutions of this system for different values of $g_{\updownarrow}$. From the real space densities $\rho_{\uparrow(\downarrow)}(\mathbf{r}) = |\Psi_{\uparrow(\downarrow)}(\mathbf{r})|^2$, we calculated the system-averaged phase separation parameter~\cite{Lin2011}, 
\begin{align}
s = \sum_{\mathbf{r}}\left[ 1 - \frac{\langle \rho_\uparrow(\mathbf{r}) \rho_\downarrow(\mathbf{r}) \rangle}{\sqrt{\langle \rho_\uparrow^2(\mathbf{r}) \rangle \langle \rho_\downarrow^2(\mathbf{r}) \rangle}}\right],
\end{align}
where the sum runs over all points $\mathbf{r} = (x,y)$ in the 2D region.
For small interspin interactions, stable vortex configurations arose with high vortex eccentricity along the y-direction~\cite{Takeuchi2020}. From the phase profile $\phi_{\uparrow(\downarrow)}(\mathbf{r})$ of each  order parameter component, we calculated the total average angular momentum, or circulation, of the pseudospinor components,
\begin{align}
    \langle \hat{L}_z\rangle_{\uparrow(\downarrow)} = \oint_{\mathcal{C}}\boldsymbol{\nabla} \phi_{\uparrow(\downarrow)}(\textbf{r}) \cdot \mathrm{d}\boldsymbol{\ell} = \frac{2\pi \hbar}{m} n_{\uparrow(\downarrow)},
\end{align}
where $\mathcal{C}$ is a closed, counter-clockwise contour enclosing 99\% of the total atom population [the thick black line in Figure \ref{fig:simresults}(d)]; the total number of $2\pi$-phase windings $n_{\uparrow(\downarrow)}$ takes on integer values since the components are single-valued. In both the phase separation and the average angular momentum characterizations, we see a clear phase transition in the ground state at $g_{\updownarrow}/g \approx 0.5$ [Figure \ref{fig:simresults}(e-f)]. 

Some obvious continuations of this work would investigate negative $g_{\updownarrow}$ values, as well as various synthetic magnetic field strengths. Field strengths are limited to a maximum value of $|B_{\uparrow, \downarrow}^*|\approx 0.700~\hbar/q a_0^2$ by the possible gauge potentials (i.e. $|\mathcal{A}^* / k_L| \in [0, 1]$) and by the physical size of the BEC. It would also be interesting to search for edge effects in a 2D uniform disk BEC~\cite{Gaunt2013, Beeler2013}.  

Throughout all this, the acceleration of our GPU method is extremely evident: a single trial typically executed in about 45 minutes on the Titan V GPU versus an estimated $\sim 1$ day on the i9 CPU. Moreover, all the results presented in Figure \ref{fig:simresults} would have taken almost 3 months of continuous computation on the i9, a highly performant device (see Table~\ref{tab:specs}).

%%%%%%%%%%%%%%%%%%%%%%%%%%%%%%%%%%%%%%%%%%%%

\begin{table*}[bt!]
    \centering
    \begin{threeparttable}
    \begin{adjustbox}{max width=\textwidth}
    \begin{tabular}{c c l l ccc}
        \hline
        Year & Source & Problem & Language/Interface & \multicolumn{3}{c}{GPU Speedup} \\
        & & & & \texttt{float32} & \texttt{float64} & \texttt{complex128} \\
        \hline
        2010 & \cite{Caplan2010} & 1D dark solitons & CUDA & 75$\times$ & 25$\times$ & - \\
        %\cline{5-6}
        2011 & \cite{Gothandaraman2011} & BEC in exiton semiconductor & CUDA & 19$\times$ & - & - \\
        %\cline{5-6}
        2013 & \cite{Caplan2013} & 1D dark solitons & MATLAB\textsuperscript{\textregistered} CUDA MEX& 37$\times$ & 31$\times$ & - \\
        %\cline{5-6}
        2013 & \cite{Demeter2013} & Non-linear optical Bloch equations & CUDA & 23$\times$ & 11$\times$ & - \\  
        2014 & \cite{Chabalko2014} & Rogue waves & CUDA & - & $>$400$\times$ & - \\  
        2015 & \cite{Berman2015} & Dipolar solitons in driven BEC & - & \multicolumn{3}{c}{10$\times^\dagger$} \\  
        2017 & \cite{Loncar2016} & Dipolar BEC & CUDA & {\color{black}-} & - & {\color{black}21 \& 25$\times$} \\  
        2016 & \cite{Kuracz2016} & Optical pulse propagation in fibers & CUDA & - & 50$\times$ & - \\  
        2020 & \cite{Gong2020} & Multimode optical fiber transmission & - & 93$\times$ & 71$\times$ & - \\  
        2021 & This work & Pseudospinor BEC & Python/PyTorch & - & - & 36 \& 85$\times$ \\  
        \hline
    \end{tabular}
    \end{adjustbox}
    \begin{tablenotes}\footnotesize
        \item[$\dagger$] Precision was not specified.
    \end{tablenotes}
    \end{threeparttable}
    \caption{Research works demonstrating GPU accelerated solutions to nonlinear Schr\"odinger equations. Almost all the works cited here were conducted in CUDA, while ours used Python. Although speedup is a relative metric, this table highlights the growing accessibility of GPU-accelerating NLSE solutions.}
    \label{tab:comparison}
\end{table*}

\section{\label{sec:Discussion}Discussion}

We have described a GPU-based approach for solving the NLSE that provided a significant speed-up, and which let us investigate details of a system that would have been otherwise inaccessible. Our computational approach was motivated by the specifics of the GPE: while some of the terms are best calculated in real space, others are better suited to momentum space, motivating FFTs between real and momentum space representations; mean field interactions and direct coupling between spinor components demanded sequential calculations.  The GPU architecture and its excellent handling of FFTs is well-suited to this algorithm.

{\color{black} For realistic modeling of experimental systems, it often becomes necessary to simulate in three dimensions. Our code is not designed for 3D, and, due to hardware memory limitations, we} note that extending {\color{black}our} procedure to 3D {\color{black} using a single GPU device} could remain quite challenging: even {\color{black}storing a complex} scalar order parameter on a cubic 3D mesh of size $1024^3$ would require over 17 GB of memory{\color{black}, exceeding} the VRAM capacity of most graphics cards on the market today.
Although better graphics cards are becoming available, solving the 3D NLSE {\color{black} with high resolution would certainly require more advanced hardware and sophisticated computational techniques than those presented here. Distributed GPU computing of the 3D NLSE has been demonstrated previously using CUDA, OpenMP, and MPI~\cite{Loncar2016b}; similar multi-device computations may be possible with PyTorch and other machine learning libraries, albeit with additional code complexity.}

{\color{black} The conversion of our existing TSSP implementation has yielded excellent performance gains; many other successful algorithms may stand to benefit from GPU acceleration in a very similar way. For instance, the previously demonstrated nonlinear conjugate-gradient technique~\cite{Gaidamour2021, Antoine2017} also relies heavily on FFTs for both the main propagation and preconditioning steps. Additionally, derivative operations can also be implemented on GPUs, so other techniques such as forward/backward-Euler finite-difference or Crank-Nicholson finite-difference~\cite{Bao2004a} may yield faster results as well. Regardless, it is also possible to pair the presented work with an existing ground-state solver~\cite{Antoine2014,  Antoine2015, Gaidamour2021, Schloss2018}; ground-state solutions found with the preferred technique can later be propagated in real-time using the presented package for faster results.}

With the availability and specifications of GPU hardware continuing to improve, we anticipate the approach taken here becoming even more widespread throughout the physics and scientific communities: without needing to know or manipulate details of the hardware architecture, one can access the advantages of GPUs while working with high-level programming tools and relatively inexpensive hardware. Though here we worked within the Python community, this approach is broadly applicable to other similarly high-level frameworks.  Within the open-source Python ecosystem, we found that many tools available through packages like Numpy and PyTorch worked well for us. We anticipate that recent updates to other packages like Numba and CuPy will offer advantages for future work in this area.

%%%%%%%%%%%%%%%%%%%%%%%%%%%%%%%%%%%%%%%%%%%
\section{\label{sec:conclusion}Conclusion}
In this paper, we demonstrated a straightforward approach to accelerate Python code solving the 2D pseudospinor nonlinear Schr\"odinger/Gross-Pitaevskii equation.
CPU-based calculations having a power-law scaling in grid size became linear by moving to a GPU-compatible system. 
We accomplished this with NVIDIA hardware upgrades, and with relatively minimal changes to our previous code, migrating from NumPy to PyTorch for the heavy calculations. Furthermore, we demonstrated their performance by simulating a spin Hall system with a spatially-varying Raman coupling. This work is a first step {\color{black} in the development process} towards hardware-accelerated code.
{\color{black} Even greater speedups are possible by rigorously optimizing the algorithm and computing resources. Our benchmarks' grid sizes were primarily limited by memory; larger grid sizes could be probed by carefully managing pre-loaded arrays and reusing temporary arrays~\cite{Loncar2016}.}
Nonetheless, the approach and \texttt{spinor-gpe} package presented in this paper illustrate the simplicity and accessibility of high-performance GPU computing for solving computationally expensive, nonlinear differential equations; these tools and methods are increasingly accessible for ``everyday'' scientific computing.
This approach is especially relevant for experimental research groups who routinely work with custom-built simulation code that is not optimized on a low level.

\section*{Competing Interests}
The authors declare that they have no known competing financial interests or personal relationships that could have appeared to influence the work reported in this paper.

\section*{Author Statement \& Acknowledgements}
Here we describe the various author contributions to this paper. 
\textit{Conceptualization}: L.J.L and L.W.C.; 
\textit{Software, Validation,  Investigation, and Formal Analysis}: L.W.C. and B.D.S.; 
% \textit{Validation}: B.D.S.; 
% \textit{Investigation}: B.D.S.;
% \textit{Formal Analysis}: B.D.S.; 
\textit{Visualization}: B.D.S.; 
\textit{Writing - Original Draft}: B.D.S. and L.W.C.; 
\textit{Writing - Review \& Editing}: B.D.S., L.W.C. and L.J.L.;
\textit{Funding Acquisition}: L.J.L, L.W.C., and B.D.S.

We wish to thank Zaheen Farraz Ahmad for many insightful discussions on GPU computing, benchmarking, and the various Python libraries available. We also gratefully acknowledge the support of NVIDIA Corporation and their grant of the Titan V GPU used in this work; and also the Natural Science and Engineering Research Council of Canada (NSERC RGPIN-2014-06618, CREATE-495446-17), Canada Foundation for Innovation (CFI), Canada Research Chairs Program (CRC), the Alberta Major Innovation Fund Quantum Technologies project, Alberta Innovates, and the University of Alberta.

%\bibliographystyle{elsarticle-num}
%\bibliography{GPUPapers}

\end{document}